\documentclass[12pt]{iopart}

\usepackage{iopams}  

\usepackage{graphicx}% Include figure files
\usepackage{dcolumn}% Align table columns on decimal point
\usepackage{bm}% bold math
\usepackage{braket}
\usepackage{color,colortbl}
\definecolor{tcA}{rgb}{0.627451,0.627451,0.643137}
\definecolor{tcC}{rgb}{0.501961,0.501961,0}
\definecolor{tcG}{rgb}{1,0,1}
\definecolor{tcH}{rgb}{0,1,1}
\definecolor{tcD}{rgb}{0.501961,0,0}
\definecolor{tcF}{rgb}{0.501961,0.501961,0.501961}
\definecolor{tcB}{rgb}{0,1,0}
\definecolor{tcE}{rgb}{1,1,0}
\newcommand{\mc}[3]{\multicolumn{#1}{#2}{#3}}
\begin{document}

\title{Resonant Population Transfer in the Time-Dependent Quantum Elliptical Billiard}

\author{F Lenz$^1$, B Liebchen$^1$, F K Diakonos$^2$ and P Schmelcher$^1$}
\address{$^1$ Zentrum f\"ur Optische Quantentechnologien, Universit\"at Hamburg,
Luruper Chaussee 149, 22761 Hamburg, Germany}
\address{$^2$ Department of Physics, University of Athens, GR-15771 Athens,
Greece}
\eads{\mailto{florian.lenz@physnet.uni-hamburg.de}, \mailto{benno.liebchen@physnet.uni-hamburg.de},  \mailto{fdiakono@phys.uoa.gr}, \mailto{peter.schmelcher@physnet.uni-hamburg.de}}

\begin{abstract}
We analyze the quantum dynamics of the time-dependent elliptical billiard using the example of a certain breathing mode. A numerical method for the time-propagation of an arbitrary initial state is developed, based on a series of transformations thereby removing the time-dependence of the boundary conditions. The time-evolution of the energies of different initial states is studied. The maximal and minimal energy that is reached during the time-evolution shows a series of resonances as a function of the applied driving frequency. At these resonances, higher (or lower) lying states are periodically populated, leading to the observed change in energy. The resonances occur when the driving  frequency or a multiple of it matches exactly the mean energetic difference between the two involved states. This picture is confirmed by a few-level Rabi-like model with periodic couplings, reproducing the key results of our numerical study.
\end{abstract}


%\submitto{\NJP}

\maketitle

\section{Introduction}
Recently, the dynamical properties of  classical time-dependent two-dimensional billiards have attracted major attention, especially in the context of Fermi acceleration \cite{Loskutov1999,Kamphorst1999,Loskutov2002,Carvalho2006,Kamphorst2007,Lenz2008, Bunimovich2010,Gelfreich2011}, which is defined as the unbounded energy gain of an ensemble of particles exposed to some external driving force \cite{Fermi1949}. Concerning the quantum dynamics of time-dependent billiards, there exist several studies investigating the quantum version of the  one-dimensional Fermi-Ulam model (or variants of it) \cite{Doescher1969,Seba1990,Makowski1991,Makowski1992,Dodonov1993,Grosche1993,Karner1994,Willemsen1994,Morales1994,Yuce2004,Jana2008,Glasser2009}, but, to our knowledge, only two studies \cite{Liboff2000,Cohen2003} investigate time-dependent billiards in higher dimensions.

Since already the one-dimensional Fermi-Ulam model involves time-dependent Dirichlet boundary conditions which are difficult to treat, both analytically and numerically, most of the works analyze  which (non-periodic) movements of the wall allow for exact solutions: Linearly expanding or contracting wall motion is considered in this context in Refs. \cite{Doescher1969} and \cite{Dodonov1993}. The authors of Refs. \cite{Makowski1991,Makowski1992} find that exact solutions are not only possible for a linear wall motion, but also for a time-law of the form $l(t)=\sqrt{at^2 +2bt+c}$ (with appropriate real constants $a,b$ and $c$). The same time-law, that allows for analytic solutions, is found from a different perspective in Refs. \cite{Grosche1993,Morales1994} and by means of a supersymmetry formalism in Ref. \cite{Jana2008}. If additionally specific electro-magnetic fields are superimposed, exact solutions can be obtained for arbitrary time-dependencies of the wall motion \cite{Yuce2004}. Periodic driving laws are considered in Refs. \cite{Seba1990} and \cite{Glasser2009}. The Floquet or quasienergy spectrum is found to be pure point like for most laws of the wall oscillation, resulting in a recurrent dynamics and thus to  bounded energy growth \cite{Seba1990}. Only  certain (non-smooth) driving laws yield a continuous quasienergy spectrum and thus allow for an unbounded acceleration, for example via resonance excitations, similarly to the quantum kicked rotator \cite{Raizen1999}. Recently, the authors of Ref. \cite{Glasser2009} studied the quantum FUM numerically by expanding the wave function in terms of the instantaneous eigenstates of the corresponding static system. They find that the dynamics of the time-dependent expansion coefficients is irregular in an intermediate frequency  regime, whereas for very low and very high frequencies a periodic behavior of the expansion coefficients is obtained.  

Concerning higher dimensional billiards, the one-pulse response of a two-dimensional stadium billiard to a deformation of the boundary has been studied \cite{Cohen2003} by analyzing the evolving energy distribution. For small wall velocities, the spreading mechanism of this distribution is dominated by transitions between neighboring levels, while this is not the case for non-adiabatic wall velocities.  In Ref. \cite{Liboff2000}, the radially vibrating three-dimensional spherical billiard is investigated. The authors claim 
\footnote{Among others, their arguments are based on the correspondence between the classical and quantum 3D driven spherical billiard. By doing so, they state that in the classical system, the angular momentum, which is a constant of the motion in the static case, gets destroyed by the driving. This is not correct, since it is known, see for example Ref. \cite{Kamphorst1999}, that in the radially oscillating two-dimensional circular billiard the angular momentum is still conserved and the arguments  can be generalized straightforwardly to the 3D spherical billiard.} that only superpositions of two or more states that share the same common rotational symmetry yield quantum chaos, since the orthogonality relations of the instantaneous eigenstates allow in any other case a reduction to a time-independent one degree of freedom Hamiltonian which cannot be chaotic. 

In the present work, we investigate the quantum version of one of the most studied classical two-dimensional time-dependent billiards, the elliptical billiard with harmonically oscillating boundaries.  Already the classical system shows several astonishing results: the driven elliptical billiard exhibits (tunable) Fermi acceleration \cite{Lenz2008} and the corresponding diffusion in momentum space experiences a dynamical crossover from sub- to normal diffusion \cite{Lenz2009,Lenz2010}. A more technical appealing aspect of using the elliptical billiard is that the quantum version of the static  billiard can be solved exactly in terms of the Mathieu functions, which allows an intuitive analysis in terms of the instantaneous eigenstates when driving the system. In particular, we will study the time-evolution of the energy in case of an initial eigenstate of the corresponding static billiard.   The  maximal and minimal energy that is reached as a function of the driving frequency show a series of resonances which we explain in detail by means of a detailed population analysis. We develop a Rabi-like model with periodic couplings which nicely reproduces the results of our full numerical simulations, confirming the intuitive picture of the level dynamics of the driven elliptical billiard.  

The work is structured as follows. In \sref{ch:static_ellise} we summarize how the eigenstates of the static elliptical billiard can be obtained by using the Mathieu functions and discuss the symmetry properties of the eigenstates. Subsequently, we tackle the problem of how to solve the driven elliptical billiard numerically in  \sref{ch:driven_ellipse}. The results of the simulations are presented in  \sref{ch:results} and interpreted, explained and modeled in \sref{ch:model}. Finally, a short summary and outlook is provided in  \sref{ch:outlook}.

\section{Static elliptical billiard}\label{ch:static_ellise}
As a precursor for the analysis of the time-dependent elliptical billiard, we briefly study the static elliptical billiard. In particular, we will use in the analysis of the time-dependent system projections on eigenstates of the static ellipse and the symmetry properties of these eigenstates.  To find the quantum mechanical eigenstates $\Psi_n(x,y)$ of a static elliptical billiard, we have to solve the corresponding stationary Schr\"odinger equation
\begin{equation}\label{eq6:schrodinger}
 H(x,y) \Psi(x,y)= E \Psi_n(x,y)
\end{equation}
where $E$ is the eigenenergy belonging to $\Psi(x,y)$ and the Hamiltonian $H(x,y)$ is given by
\begin{equation}\label{eq6:hamiltonian}
 H(x,y)= -\frac{\hbar^2}{2\mu}\left (\frac{\partial^2}{\partial x^2}+\frac{\partial^2}{\partial y^2}\right ) +V(x,y),
\end{equation}
where $\mu$ denotes the mass. The potential $V(x,y)$ is zero inside and infinite outside the static elliptical billiard
\begin{equation}\label{eq6:potential}
 V(x,y)=\left\{
 \begin{array}{ll}
  0 & \rm{if } \frac{x^2}{a^2} +\frac{y^2}{b^2}\leq 1\\
   \infty & \rm{if } \frac{x^2}{a^2} +\frac{y^2}{b^2}>1.
 \end{array} \right.
\end{equation}
This means the eigenvalue equation \eref{eq6:schrodinger} has to be solved under Dirichlet boundary conditions
\begin{equation}\label{eq6:boundary_cond}
 \Psi(x,y)=0 \quad \mbox{for } \left \{(x,y)^\top \in \mathbb R^2\bigg \vert \frac{x^2}{a^2}+\frac{y^2}{b^2}=1 \right\},
\end{equation}
where $a$ and $b$ are the semi-major and semi-minor axis of the ellipse, respectively.

Due to the boundary condition of Eq.~\eref{eq6:boundary_cond}, solving the eigenvalue equation~\eref{eq6:schrodinger} is actually not trivial and involves some subtleties \cite{Zon1998,Traiber1989,Ayant1987,Arvieu1987,Waalkens1997,Vega1999,Chavez2001}. Here, we follow the discussion provided in Ref.~\cite{Chavez2001} and additionally use results presented in Refs. \cite{Vega1999} and \cite{Waalkens1997}. We start by transforming from Cartesian to elliptic coordinates:
\numparts
 \begin{eqnarray}\label{eq6:elliptic_coord}
 x&= f \cosh \xi \cos \eta, \quad 0 \leq \xi \leq \xi_0\\
 y&= f \sinh \xi \sin \eta, \quad 0 \leq \eta < 2\pi,
\end{eqnarray}
\endnumparts
where $f=\sqrt{a^2-b^2}$ is the semi-focal distance of the ellipse and $\xi_0={\rm{arctanh}}(b/a)$. The stationary Schr\"odinger equation reads
\begin{equation}\label{eq6:sd}
  H(\xi, \eta)\varphi(\xi,\eta)= E \varphi(\xi,\eta),
\end{equation}
where $\varphi(\xi,\eta)$ is an eigenfunction with energy $E$ and $H(\xi, \eta)$ is the Hamiltonian:
\begin{equation}
 H(\xi, \eta)= -\frac{\hbar^2}{2\mu} \bigtriangleup(\xi,\eta) + V(\xi, \eta)
\end{equation}
The potential energy $V(\xi, \eta)$  in elliptic coordinates is now
\begin{equation*}
V(\xi, \eta) = \left\{
 \begin{array}{ll}
  0 & \rm{if } \xi \leq  \xi_0\\
   \infty & \rm{if } \xi > \xi_0,
 \end{array} \right.
\end{equation*}
As a consequence, the eigenfunctions $\varphi(\xi,\eta)$ have to satisfy the Dirichlet boundary condition $\varphi(\xi_0,\eta)=0$ and also the periodicity condition $\varphi(\xi,\eta)=\varphi(\xi,\eta+2\pi)$. The Laplacian operator in elliptic coordinates is given by
\begin{equation}
 \bigtriangleup(\xi,\eta)=\frac{2}{f^2(\cosh 2\xi -\cos 2\eta)}\left (\frac{\partial^2}{\partial \xi^2}+\frac{\partial^2}{\partial \eta^2} \right).
\end{equation}
The two-dimensional (2D) Schr\"odinger equation inside the elliptical billiard (i.e. $\xi\leq\xi_0$) then simplifies to
\begin{equation}\label{eq:2d_se}
 \left [\frac{\partial^2}{\partial \xi^2}+\frac{\partial^2}{\partial \eta^2}  + \frac{k^2 f^2}{2}(\cosh 2\xi -\cos 2\eta)\right]\varphi(\xi,\eta)=0,
\end{equation}
with $k^2=2\mu E/\hbar^2$. Obviously, $\varphi(\xi,\eta)$ can be separated with the standard ansatz
\begin{equation}\label{eq6:ansatz}
 \varphi(\xi,\eta)=R(\xi)\Theta(\eta)
\end{equation}
into a radial ($R(\xi)$) and an angular ($\Theta(\eta)$) part. Plugging this separation ansatz in the Schr\"odinger equation \eref{eq:2d_se}, we obtain two ordinary differential equations:
\numparts\label{eq6:mathieu}
 \begin{eqnarray}\label{eq:math1}
 R''(\xi)-(\alpha-2q \cosh 2\xi)R(\xi)&=&0,\\
\Theta''(\eta)+(\alpha-2q \cos 2\eta)\Theta(\eta)&=&0,\label{eq:math2}
\end{eqnarray}
\endnumparts
where $\alpha$ is the separation constant and $q$ is the dimensionless, rescaled energy 
\begin{equation}\label{eq6:separation_q}
 q=\frac{k^2 f^2}{4}=\frac{\mu f^2}{2\hbar^2} E. 
\end{equation}
Equations \eref{eq:math1}, \eref{eq:math2} are the standard form of the Mathieu equations, Eq. \eref{eq:math2} is the ordinary Mathieu equation (OME) and Eq. \eref{eq:math1} is the modified Mathieu equation (MME). The corresponding solutions are the ordinary and modified Mathieu functions (OMF, MMF), respectively. Note that the change of variables $\eta=i\xi$ transforms the OME into the MME. Even though the ansatz \eref{eq6:ansatz} decouples the Schr\"odinger equation \eref{eq:2d_se}, yielding  two ordinary differential equations, the separation constants $\alpha$ and $q$ do not decouple \cite{Morse1953}, as we will see in the following.

The physical solutions of the OME have to be periodic, i.e. they can be expanded in a Fourier series (actually, Floquet theory guarantees the existence of periodic solutions \cite{Waalkens1997})
\begin{equation}\label{eq6:solution_ome}
\Theta_l(\eta)= 
\left\{
 \begin{array}{ll}
  {\rm ce}_{l}(\eta,q)= \sum_{k=0}^{\infty} A_k(l,q) \cos (k\eta)\\
  {\rm se}_{l}(\eta,q)= \sum_{k=1}^{\infty} B_k(l,q) \sin (k\eta)
 \end{array} \right.
\end{equation}
For a fixed $q$, it is known that only certain values of $\alpha$ allow for periodic solutions. These eigenvalues are called the characteristic values $\alpha_l$ and $\beta_{l}$, where $l$ is the order ($l\geq 0$ for ${\rm ce}_l(\eta)$ and $l\geq 1$ for ${\rm se}_l(\eta)$). Since the Mathieu equation is of Sturm-Liouville type, all eigenvalues are real, positive and can be ordered $\alpha_0 < \beta_1 < \alpha_1< \beta_2 \dots$. In the $q\times \alpha$ plane, the functions $\alpha_l(q)$ and $\beta_{l}(q)$ are curves that do not intersect. The  ${\rm ce}_{l}(\eta,q)$ and ${\rm se}_{l}(\eta,q)$ with even order have a period of $\pi$, the ones with odd order have a period of $2\pi$. The order $l$ specifies the number of zeros of ${\rm ce}_{l}(\eta,q)$ and ${\rm se}_{l}(\eta,q)$ in the interval $\eta \in [0,\pi]$. The expansion coefficients $A_k(l,q)$ and $B_k(l,q)$ are determined by the recurrence relations for the Mathieu equations \cite{Abramowitz1972}.

The solutions of the MME Eq. \eref{eq:math1}, i.e. the radial part, can be obtained from the solutions of the OME by the mentioned change of variable $\eta= i\xi$, yielding
\begin{equation}\label{eq6:solution_mme}
R_l(\xi)= 
\left\{
 \begin{array}{ll}
  {\rm Ce}_{l}(\xi,q)= \sum_{k=0}^{\infty} A_k(l,q) \cosh (k\xi)\\
  {\rm Se}_{l}(\xi,q)= \sum_{k=1}^{\infty} B_k(l,q) \sinh (k\xi).
 \end{array} \right.
\end{equation}
Note that the expansion coefficients $A_k(l,q)$ and $B_k(l,q)$ are the same as the ones in the solutions \eref{eq6:solution_ome}
of the OME.  The solutions of the MME can also be viewed as an independent eigenvalue problem for the $q$'s (with fixed $\alpha$ and $\beta$), now $R_l(\xi)$ has to the satisfy the Dirichlet boundary condition $R_l(\xi_0)=0$.

The eigenvalues associated to $\alpha$ and $\beta$ are denoted by $q_r^e(\alpha)$ and $q_r^o(\beta)$, where the index $o,e$ refers to the even and odd MMF and $r\geq 1$ is the $r$th zero of the MMF, not counting a possible root at $\xi=0$. In the $q\times \alpha$ plane, the functions $q_r^e(\alpha)$ and $q_r^o(\beta)$ are curves that again do not intersect. The solution $\varphi(\xi,\eta)$ of the Schr\"odinger equation \eref{eq:2d_se} is the product \eref{eq6:ansatz} of the OMF and MMF. Both eigenvalue problems have to be solved simultaneously, since $\alpha$ (or $\beta$) and $q^e$ (or $q^o$) in the OMF and MMF of the solution $\varphi(\xi,\eta)$ cannot be chosen independently. The pairs $(\alpha, q)$ (or $(\beta, q)$) that satisfy both eigenvalue problems simultaneously are of course the crossing points of the family of curves $\alpha_l(q)$ and $\beta_{l}(q)$ with the $q_r^e(\alpha)$ and $q_r^o(\beta)$. Thus, there are two types of crossing points that correspond to solutions of the simultaneous eigenvalues problem:
\begin{enumerate}
 \item $\alpha_r(q)$ with $q_m^e(\alpha)$ (even modes)
\item $\beta_{r}(q)$ with $q_m^o(\beta)$ (odd modes)
\end{enumerate}
These two types of crossing points are related to the symmetries of the solution $\varphi(\xi,\eta)$. This yields the  eigenfunctions $\varphi_{l,r}(\xi,\eta)$ 
\numparts
 \begin{eqnarray}
  \varphi_{l,r}^e(\xi,\eta)&= {\rm Ce}_l(\xi,q^e_{l,r}){\rm ce}_l(\eta,q^e_{l,r})\\
  \varphi_{l,r}^o(\xi,\eta)&= {\rm Se}_l(\xi,q^o_{l,r}){\rm se}_l(\eta,q^o_{l,r}),
 \end{eqnarray}
\endnumparts
where $q^e_{l,r}=q^e_r(\alpha_l)$ and correspondingly  $q^o_{l,r}=q^o_r(\beta_l)$. The indices $l$ and $r$ are the angular and radial quantum numbers of the even and odd modes, respectively. $l$ is the order of the OMF  and $r$ specifies the number of zeros of the MMF in the interval $\xi\in [0,\xi_0]$, not counting the zeros at the origin  i.e. $q^{e,o}_{l,r}$ is the $r$-th zero of the MMF of order $l$. Since the MMF has to be zero at $\xi_0$, it follows that $r$ is greater or equal to one. According to Eq. \eref{eq6:separation_q}, the separation constant $q$ is directly proportional to the energy and the energy eigenvalues $E^{e,o}_{l,r}$ corresponding to the eigenfunctions $\varphi_{l,r}^{e,o}(\xi,\eta)$ can be written as
\begin{equation}\label{eq:energy_epsilon}
 E^{e,o}_{l,r}= \frac{2\hbar^2}{\mu f^2}q^{e,o}_{l,r}= \frac{2\hbar^2}{\mu a^2 \epsilon^2}q^{e,o}_{l,r},
\end{equation}
where $\epsilon$ is the eccentricity of the ellipse.

\subsection*{Symmetries}
The symmetries of the eigenstates $\Psi^{e,o}_{l,r}(x,y)$ of the elliptical billiard are determined by the OMF $\Theta_l(\eta)$ \eref{eq6:solution_ome} \cite{Waalkens1997}. For an even OMF, i.e. for ${\rm ce}_l(\eta)$, the wavefunction $\Psi^{e}_{l,r}(x,y)$ is symmetric with respect to reflections at the $x$-axis ($\Psi^{e}_{l,r}(x,y)=\Psi^{e}_{l,r}(x,-y)$), this symmetry is denoted by $\pi_y=+1$. Odd OMF ${\rm se}_{l+1}(\eta)$ imply an antisymmetric eigenstate  $\Psi^{o}_{l,r}(x,y)$ with respect to the $x$-axis ($\Psi^{e}_{l,r}(x,y)=-\Psi^{e}_{l,r}(x,-y)$), denoted by $\pi_y=-1$. The symmetry of $\Psi^{e,o}_{l,r}(x,y)$ with respect to the $y$-axis is fixed by the angular quantum number $l$, which is the order of the OMF. For even Mathieu functions, $\Psi^{e}_{l,r}(x,y)$ is symmetric with respect to reflections at the $y$-axis ($\Psi^{o}_{l,r}(x,y)=\Psi^{o}_{l,r}(-x,y)$) if $l$ is even and antisymmetric ($\Psi^{o}_{l,r}(x,y)=-\Psi^{o}_{l,r}(-x,y)$) if $l$ is odd. Naturally, this symmetry is denoted by $\pi_x$. For odd  Mathieu functions, $\Psi^{o}_{l,r}(x,y)$ is symmetric to the $y$-axis if $l$ is odd and  antisymmetric if $l$ is even. Overall, this yields four possible combinations concerning the  parity of the eigenstates $\Psi^{e,o}_{l,r}(x,y)$, which are summarized in the following list:
\begin{enumerate}
 \item $\pi_y=+1, \,\pi_x=+1$ $\Rightarrow$ $\Psi^{e}_{l,r}(x,y)=\Psi^{e}_{l,r}(-x,-y)$ even solutions with even $l$
\item $\pi_y=+1, \,\pi_x=-1$ $\Rightarrow$ $\Psi^{e}_{l,r}(x,y)=-\Psi^{e}_{l,r}(-x,-y)$ even solutions with odd $l$
\item $\pi_y=-1, \,\pi_x=+1$ $\Rightarrow$ $\Psi^{o}_{l,r}(x,y)=-\Psi^{o}_{l,r}(-x,-y)$ odd solutions with odd $l$
\item $\pi_y=-1, \,\pi_x=-1$ $\Rightarrow$ $\Psi^{o}_{l,r}(x,y)=\Psi^{o}_{l,r}(-x,-y)$ odd solutions with even $l$
\end{enumerate}
The four parity combinations correspond to the characterization of the symmetry reduced quarter elliptic billiard. However, the boundary conditions along the coordinate axes of the quarter billiard have to be adjusted according to the different parities: Dirichlet boundary conditions are required along the $x$-axis ($y$-axis) for $\pi_y=-1$ ($\pi_x=-1$) and Neumann boundary conditions along the $x$-axis ($y$-axis) for $\pi_y=+1$ ($\pi_x=+1$). 

\begin{table}
\caption{All eigenstates $\Psi^{e,o}_{l,r}(x,y)$ of the elliptical billiard ($a=1$, $b=\sqrt{0.51}\,\Rightarrow$ eccentricity $\epsilon=0.7$) with $E<50$ (in units of $\hbar^2/\mu$). The eigenstates are ordered according to the energy. $\pi_y$ ($\pi_x$) denotes the symmetry of $\Psi^{e,o}_{l,r}(x,y)$ with respect to reflections at the $x$-axis ($y$-axis).}
\begin{center}
 \begin{tabular}{l|c|c|c|c|c|c}\label{tb:tab1}
Label & $E [\hbar^2/\mu]$ & $\,l\,$ & $\,r\,$ & even/odd & $\,\pi_y\,$ & $\,\pi_x\,$\\
\hline
1&	4,267&	0&	1&	e& $+$ & $+$\\
2&	9,058&	1&	1&	e& $+$ & $-$\\
3&	12,577&	1&	1&	o& $-$ & $+$\\
4&	15,993&	2&	1&	e& $+$ & $+$\\
5&	19,358&	2&	1&	o& $-$ & $-$\\
6&	25,061&	3&	1&	e& $+$ & $-$\\
7&	25,895&	0&	2&	e& $+$ & $+$\\
8&	27,998&	3&	1&	o& $-$ & $+$\\
9&	35,156&	1&	2&	e& $+$ & $-$\\
10&	36,178&	4&	1&	e& $+$ & $+$\\
11&	38,511&	4&	1&	o& $-$ & $-$\\
12&	44,040&	1&	2&	o& $-$ & $+$\\
13&	46,406&	2&	2&	e& $+$ & $+$\\
14&	49,199&	5&	1&	e& $+$ & $-$
\end{tabular}
\end{center}
\end{table}

\section{Driven elliptical billiard}\label{ch:driven_ellipse}
To study the time evolution of an initial state $\Psi_0(x,y):=\Psi(x,y,t=0)$  in the harmonically oscillating elliptical billiard, we have to solve the time-dependent Schr\"odinger equation 
\begin{equation}\label{eq7:schrödinger}
 i\hbar \frac{\partial \Psi (x,y,t)}{\partial t} = H(x,y,t) \Psi(x,y,t),
\end{equation}
where the Hamiltonian $H(x,y,t)$ is given by
\begin{equation}\label{eq7:hamiltonian}
 H(x,y,t)= -\frac{\hbar^2}{2\mu}\left (\frac{\partial^2}{\partial x^2}+\frac{\partial^2}{\partial y^2}\right ) +V(x,y,t)
\end{equation}
and the potential $V(x,y,t)$ is zero inside and infinity outside the time-dependent elliptical billiard:
\begin{equation}\label{eq7:td_potential}
 V(x,y,t)=\left\{
 \begin{array}{ll}
  0 & \rm{if } \frac{x^2}{a^2(t)} +\frac{y^2}{b^2(t)}\leq 1\\
   \infty & \rm{if } \frac{x^2}{a^2(t)} +\frac{y^2}{b^2(t)}>1.
 \end{array} \right.
\end{equation}
This means  the Schr\"odinger equation \eref{eq7:schrödinger} has to be solved under Dirichlet boundary conditions
\begin{equation}
 \Psi(x,y,t)=0 \quad \mbox{for } \left \{(x,y)^\top \in \mathbb R^2\bigg \vert \frac{x^2}{a^2(t)}+\frac{y^2}{b^2(t)}=1,\, \forall t \right\},
\end{equation}
where $a(t)$ and $b(t)$ are determined by the driving law of the elliptical billiard.

\subsection*{Transformations}
The explicit  time-dependence of the boundary conditions is very difficult to be treated by standard numerical techniques. Therefore, to remove the explicit time-dependence of the boundary conditions, we apply the following time-dependent coordinate transformation, whose 1D variant has been successfully used to remove time-dependent boundary conditions \cite{Doescher1969,Makowski1991,Makowski1992,Dodonov1993,Morales1994,Yuce2004,Jana2008,Glasser2009} in 1D systems
\begin{equation}\label{eq7:coord_trans}
 \eta = \frac{x}{a(t)}, \quad \xi=\frac{y}{b(t)},
\end{equation}
yielding the time-dependent Schr\"odinger equation
\begin{equation}\label{eq7:schrodinger_eta_xi}
 i\hbar \frac{\partial \Psi (\eta(t),\xi(t),t)}{\partial t} = H(\eta(t),\xi(t),t) \Psi(\eta(t),\xi(t),t).
\end{equation}
The advantage of applying this time-dependent coordinate
transformation is that the boundary condition in the new
coordinates $\eta,\xi$ is extremely simple: the wave function
$\Psi(\eta,\xi,t)$ has to vanish on the circle given by
$\eta^2+\xi^2=1$. The prize we have to pay is that now the
coordinates themselves are explicitly time-dependent, $\eta=\eta(t)$ and
$\xi=\xi(t)$. This has to be taken into account when
applying the differential operator $\partial /\partial t$ on
$\Psi(\eta(t),\xi(t),t)$ in Eq.~\eref{eq7:schrodinger_eta_xi}. The additional terms resulting from the left hand side of Eq.~\eref{eq7:schrodinger_eta_xi} can be put into an effective Hamiltonian, yielding 
\begin{eqnarray}\label{eq7:ham2}
 H^e(\eta, \xi, t)= -\frac{\hbar^2}{2\mu}\left( \frac{1}{a^2(t)}\frac{\partial^2}{\partial \eta^2}+\frac{1}{b^2(t)}\frac{\partial^2}{\partial \xi^2}\right )\nonumber \\ + i\hbar \left ( \frac{\dot{a}(t)}{a(t)}\eta\frac{\partial}{\partial \eta}+\frac{\dot{b}(t)}{b(t)}\xi\frac{\partial}{\partial \xi}\right).
\end{eqnarray}
and the corresponding time-dependent Schr\"odinger equation
\begin{equation}\label{eq7:schrodinger2}
 i\hbar \frac{\partial \Psi (\eta,\xi,t)}{\partial t}\bigg \vert_{\eta,\xi=const.} = H^e(\eta,\xi,t) \Psi(\eta,\xi,t).
\end{equation}
To  remove the terms proportional to $i\hbar \eta\frac{\partial}{\partial \eta}$
and $i\hbar \xi\frac{\partial}{\partial \xi}$, we apply the following
time-dependent  unitary transformation, yielding the new wave function $\Lambda(\eta,\xi,t)$
\begin{eqnarray}\label{eq7:unitary_transformation}
 \Psi(\eta, \xi, t) &= \Omega(\eta, t)\cdot \Upsilon(\xi,t)\cdot \Lambda(\eta, \xi,t)\\
\Psi(\eta, \xi,t) &= \frac{2}{\sqrt{ab}}\exp \left(
\frac{i\mu}{2\hbar}(\dot aa\eta^2 +\dot bb\xi^2)\right)\cdot
\Lambda(\eta, \xi,t).
\end{eqnarray}
With this transformation we obtain the Schr\"odinger equation
\begin{equation}\label{eq7:schodinger_xi_n}
 i\hbar \frac{\partial \Lambda (\eta,\xi,t)}{\partial t} = \mathcal H^e(\eta,\xi,t) \Lambda(\eta,\xi,t),
\end{equation}
with the effective Hamiltonian
\begin{eqnarray}\label{eq7:ham_hermitian}
 \mathcal H^e(\eta,\xi,t)= -\frac{\hbar^2}{2\mu}\left( \frac{1}{a^2(t)}\frac{\partial^2}{\partial \eta^2}+\frac{1}{b^2(t)}\frac{\partial^2}{\partial \xi^2}\right )\nonumber \\ +\frac12 \mu a(t)\ddot a(t)\eta^2 +\frac12 \mu b(t)\ddot b(t)\xi^2
\end{eqnarray}
and the time-independent boundary condition $\Lambda(\eta,\xi,t)=0$ for $\eta^2+\xi^2=1$. This effective Hamiltonian can be interpreted as a two-dimensional anisotropic harmonic oscillator with
time-dependent frequencies and time-dependent masses and the above boundary condition.

We expand $\Lambda(\eta,\xi,t)$ in terms of the eigenfunctions $\Phi_{n,m}$ of the (unit) static circular billiard, since this ansatz automatically fulfills the boundary condition. The eigenfunctions $\Phi_{n,m}$ of the static circular billiard are best described in polar coordinates, so we transform $\eta$ and $\xi$ to polar coordinates:
\begin{equation}\label{eq7:polar_transform1}
 \eta=r\cos \phi \quad \xi=r\sin \phi \quad \Rightarrow \Lambda(r=1, \phi, t)=0, \,\forall \phi,t
\end{equation}
The Hamiltonian \eref{eq7:ham_hermitian} in polar coordinates reads
\begin{eqnarray}\label{eq7:hamiltonian_polar}
\mathcal H^e(r,\phi,t)= -\frac{\hbar^2}{2\mu a^2(t)} \bigg(\sin^2 \phi \frac{\partial^2}{\partial r^2} -\frac{2\sin \phi \cos \phi}{r^2} \frac{\partial}{\partial \phi}\nonumber \\
+\frac{2\sin \phi \cos \phi}{r} \frac{\partial^2}{\partial r\partial \phi} 
+\frac{\cos^2 \phi}{r} \frac{\partial}{\partial r} + \frac{\cos^2
\phi}{r^2} \frac{\partial^2}{\partial \phi^2}\bigg)\nonumber \\
-\frac{\hbar^2}{2\mu b^2(t)} \bigg(\cos^2 \phi
\frac{\partial^2}{\partial r^2} +\frac{2\sin \phi \cos \phi}{r^2}
\frac{\partial}{\partial \phi} -\frac{2\sin \phi \cos \phi}{r}
\frac{\partial^2}{\partial r\partial \phi}\nonumber \\ +\frac{\sin^2 \phi}{r}
\frac{\partial}{\partial r} + \frac{\sin^2 \phi}{r^2}
\frac{\partial^2}{\partial \phi^2}\bigg)  +\frac12 \mu a(t)\ddot
a(t) r^2 \cos^2 \phi \nonumber \\ +\frac12 \mu b(t)\ddot b(t)r^2 \sin^2 \phi
\end{eqnarray}
and the corresponding Schr\"odinger equation is 
\begin{equation}\label{eq7:schrodinger_polar}
 i\hbar \frac{\partial \Lambda (r,\phi,t)}{\partial t} = \mathcal H^e(r,\phi,t) \Lambda(r,\phi,t).
\end{equation}
\subsection*{Ansatz}
As already written, we expand the wave function in terms of the eigenfunctions $\Phi_{n,m}(r,\phi)$ of the static circular billiard. To carry out numerical simulations, this expansion has to be truncated, yielding
\begin{equation}\label{eq7:ansatz}
 \Lambda(r,\phi,t)=\sum_{n=1}^{N} \sum_{m=-M}^{M} c_{n,m}(t)\Phi_{n,m}(r,\phi),
\end{equation}
with time-dependent expansion coefficients $c_{n,m}(t)$, where $n$ is the radial and $m$ the azimuthal quantum number. The time-independent eigenfunctions $\Phi_{n,m}(r,\phi)$ of the static circular billiard factorize
\begin{equation}\label{eq7:factorize}
 \Phi_{n,m}(r,\phi)= R_{n,m}(r)\cdot \Theta_m(\phi).
\end{equation}
The normalized radial and azimuthal (angular) functions are given by \cite{Robinett1996,Robinett2003}
\begin{eqnarray}\label{eq7:eigenf_circle}
 R_{n, m }(r) &= \frac{\sqrt{2}}{J_{ m  +1}(k_{n, m})} J_{ m }(k_{m, n}r) \\
\Theta_m(\phi)&=\frac{1}{\sqrt{2\pi}}e^{im\phi}
\end{eqnarray}
where $J_{ m}$ is the Bessel function of the first kind of order $m$ and $k_{m,n}$ is the $n$th zero of $J_{m}$. 

To determine the time-dependent expansion coefficients $c_{n,m}(t)$, we insert the ansatz \eref{eq7:ansatz} into the time-dependent Schr\"odinger equation \eref{eq7:schrodinger_polar}, project with the bra $\bra{ \Lambda}$
\begin{equation}\label{eq7:PsiHPsi}
 \bra{\Lambda}i\hbar \frac{\partial}{\partial t}\ket{\Lambda}= \bra{\Lambda}\mathcal  H^e(t) \ket{\Lambda}
\end{equation}
and thus  obtain a set of coupled ordinary differential equations (ODE) for the $c_{n,m}(t)$. This coupled ODE system can be written in the following form, for details of the derivation see Ref. \cite{LenzThesis}
\begin{eqnarray}\label{eq7:ode_short}
 i\hbar \dot c_{n,m}(t)= \sum_{n'}\big \{c_{n',m}(t)d^{(1)}_{nmn'}(t)\nonumber \\+c_{n',m-2}(t)d^{(2)}_{nmn'}(t)+c_{n',m+2}(t)d^{(3)}_{nmn'}(t)\big \}.
\end{eqnarray} 
The $d^{(i)}_{nmn'}(t)$ can be written in the following form (exemplarily shown for $d^{(1)}_{nmn'}(t)$)
\begin{equation*}
 d^{(1)}_{nmn'}(t)= g^{(1)}(t)f^{(1)}_{nmn'}+g^{(1)}(t)f^{(2)}_{nmn'},
\end{equation*}
where the $g^{(i)}(t)$ are simple time-dependent functions of $a(t)$  and $b(t)$, whereas the $f^{(i)}_{nmn'}$ are time-independent. For a definition of the $d^{(i)}_{nmn'}(t)$, the $f^{(i)}_{nmn'}$ and the $g^{(i)}(t)$, see  \ref{ch:appendix}. More specific, the $f^{(i)}_{nmn'}$ are composed of integrals over products of Bessel functions, which in general cannot be evaluated analytically.  However, since the corresponding integrals are time-independent, they have to be evaluated only once, which can be done numerically with high precision. Thus, by introducing a linear index $(n,m)\leftrightarrow i$ the coupled ODE system~\eref{eq7:ode_short} is in the canonical form and can be solved numerically by standard techniques.  

\subsection*{Observables}
The time-evolution of  the expectation value of the  energy of a certain initial state propagating in the driven elliptical   is given by
\begin{equation}\label{eq:energy1}
 E(t)= \bra{ \Psi(t)} H(t)\ket{\Psi(t)}.
\end{equation}
To calculate $E(t)$ at a given time $t_0$ of the original laboratory system, we cannot directly use the wavefunction $\Lambda(r,\phi,t_0)$ \eref{eq7:ansatz} together with the Hamiltonian $\mathcal H^e(r,\phi,t_0)$ \eref{eq7:hamiltonian_polar}.  Instead we have to apply the following steps: First, the inverse of the  transformations \eref{eq7:coord_trans}, \eref{eq7:unitary_transformation} and \eref{eq7:polar_transform1} operate on $\Lambda(r,\phi,t_0)$ and we obtain
\begin{eqnarray}\label{eq7:psi_transformed}
 \Psi(x,y,t_0)= \frac{1}{\sqrt{ab}} e^{\frac{i\mu }{2 \hbar}\left( \dot a|_{t=t_0}x^2/a(t_0)+\dot b|_{t=t_0}y^2/b(t_0) \right)}\times  \nonumber \\ \sum_{n,m} c_{n,m}(t_0)\Phi_{n,m}\left(\sqrt{\frac{x^2}{a^2}+\frac{y^2}{b^2}},\arctan\left(\frac{ay}{bx}\right)\right).
\end{eqnarray}
Secondl, we apply the  coordinate transformation as given in Eq. \eref{eq7:coord_trans}, but now as a time-independent coordinate transformation, which just depends parametrically on the time $t_0$: 
\begin{equation}\label{eq7:coord_trans_s}
 \eta = \frac{x}{a(t_0)}, \quad \xi=\frac{y}{b(t_0)} 
\end{equation}
The wavefunction and the Hamiltonian are then given by
\begin{eqnarray}
 \Psi(\eta,\xi,t_0)= 2 e^{\frac{i\mu }{2 \hbar}\left( \dot a|_{t=t_0}a(t_0) \eta^2+\dot b|_{t=t_0}b(t_0) \xi^2 \right)}\times \nonumber  \\
\sum_{n,m} c_{n,m}(t_0)\Phi_{n,m}\left(\sqrt{\eta^2+\xi^2},\arctan\left(\frac{\xi}{\eta}\right)\right),
\end{eqnarray}
\begin{equation}\label{eq:ham_tid1}
 H(\eta, \xi, t_0)= -\frac{\hbar^2}{2\mu }\left( \frac{1}{a^2(t_0)}\frac{\partial^2}{\partial \eta^2}+\frac{1}{b^2(t_0)}\frac{\partial^2}{\partial \xi^2}\right ).
\end{equation}
Finally, changing again to polar coordinates yields
\begin{equation}\label{eq7:psi_prefactor}
 \Psi(r,\phi,t_0)= e^{\frac{i\mu }{2 \hbar}r^2\left( \dot aa +(\dot bb -\dot aa)\sin^2 \phi  \right)} \sum_{n,m} c_{n,m}(t_0)\Phi_{n,m}(r,\phi),
\end{equation}
and the corresponding Hamiltonian $\mathcal H(r,\phi,t)$ is given by Eq.~\eref{eq7:hamiltonian_polar} if we set $\ddot a(t)=\ddot b(t)=0$, i.e. 
\begin{equation}
 \mathcal H(r,\phi,t)= \mathcal H^e(r,\phi,t)\big \vert_{\ddot a=\ddot b=0}.
\end{equation}
Another quantity of interest is the spectral composition of the energy in terms of the population of the  instantaneous eigenstates of the elliptical billiard. By `instantaneous eigenstates' we mean the following: Consider the boundary of the elliptical billiard for a certain time instant. If we now take this particular billiard configuration, treat it as a static ellipse and calculate the corresponding eigenstates assuming Dirichlet boundary conditions, we obtain the  instantaneous eigenstates. If we denote the eigenstates by $\bra{\Psi_i(t)}$ ($i$ labels the degree of energetic excitation), the population $p_i$ of an instantaneous eigenstate is then of course simply the projection of the wave function $\Psi(x,y,t)$ onto this state, i.e. $p_i(t)=|\braket{\Psi_i(t)|\Psi(t)}|$.

\begin{figure}%[ht]
\centerline{\includegraphics[width=0.7\columnwidth]{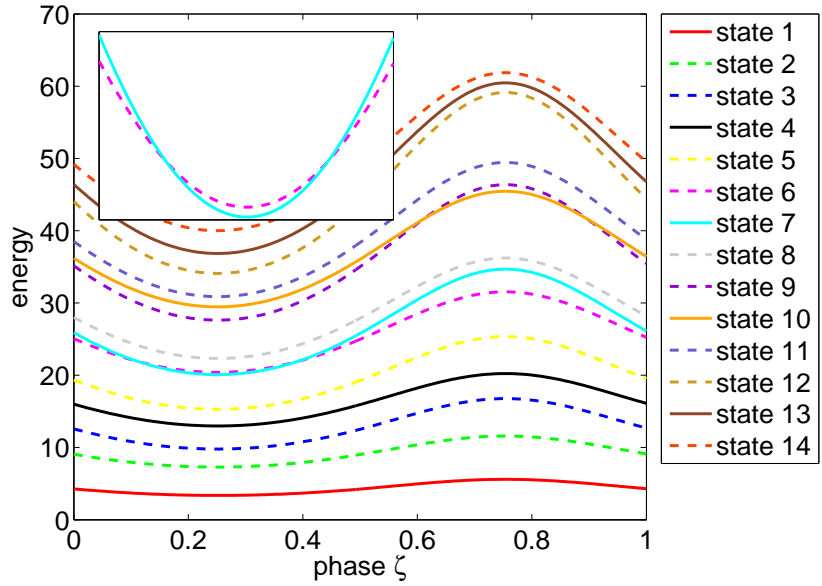}}
\caption{(Color online) Energy $E_i(\zeta)$  (in units of $\hbar^2/\mu$, the units will be omitted in the following for simplicity) of the instantaneous eigenstates as a function of the phase $\zeta$ of the elliptical billiard.  The solid lines represent the energetically lowest  five eigenstates with symmetries $\pi_x=\pi_y=+1$.  The inset shows the crossings of the states six and seven.\label{fig:fig1}}
\end{figure}

It is illustrative to analyze how the energy of the instantaneous eigenstates changes during one oscillation of the elliptical billiard. From Eq.~\eref{eq:energy_epsilon} it can be seen that the energy of the eigenstates of the elliptical billiard depends (for a given semi-major axis $a$) on the eccentricity $\epsilon$, so $E_i=E_i(\epsilon)$. Since the periodic (with period $T=2\pi/\omega$) driving law $a(t)$, $b(t)$ changes the eccentricity, we may write for the energy of the instantaneous eigenstates $E_i(\zeta)$, where $\zeta$ is the phase of the oscillation, measured in terms of $T$, i.e. $\zeta \in [0, 1]$. $E_i(\zeta)$ for energetically lowest 14 states is shown in \fref{fig:fig1}.  There are several crossings, note that these are real crossings and not avoided crossings, see the inset. Such crossings are possible since the static elliptical billiard is an integrable system, thus the corresponding eigenstates posses `good' quantum numbers, in agreement with the theorem of Wigner and von Neumann \cite{Wigner1929}. 
\begin{figure}%[ht]
\centerline{\includegraphics[width=0.7\columnwidth]{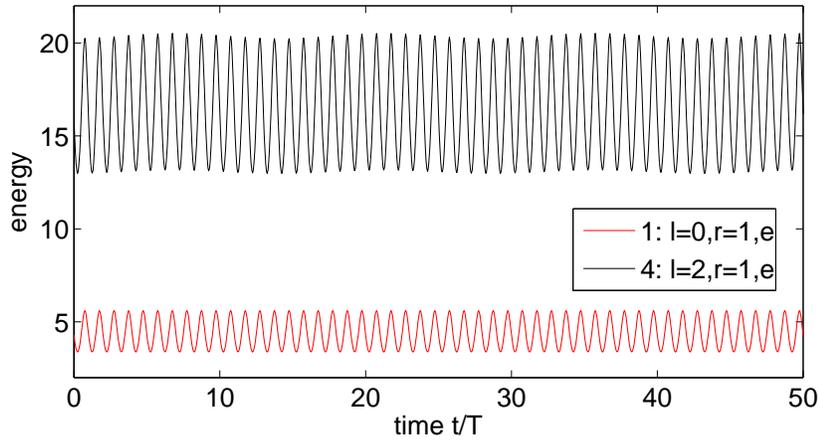}}
\caption{(Color online) Evolution of the energy $ E(t)$ for a driving frequency of $\omega=5$, when starting in the instantaneous ground state (first state) and in the fourth state. \label{fig:fig2}}
\end{figure}

\section{Results}\label{ch:results}  
In the following, we will take as an initial state $\Psi_0=\Psi(t=0)$ an eigenstate of the static elliptical billiard, i.e. $\Psi_0=\Psi^{e,o}_{l,r}(x,y)$, and let it evolve in the driven ellipse. We choose for the driving law of the elliptical billiard $a(t)=a_0+c\sin(\omega t)$ and $b(t)=b_0+c\sin(\omega t)$. For the equilibrium positions of the semi-major and semi-minor axes we use $a_0=1$, $b_0=\sqrt{0.51}$, resulting in an eccentricity of $\epsilon_0=0.7$. The driving amplitude $c$ is set to $0.1$, whereas the frequency $\omega$ will be varied.

\begin{figure}%[ht]
\centerline{\includegraphics[width=0.7\columnwidth]{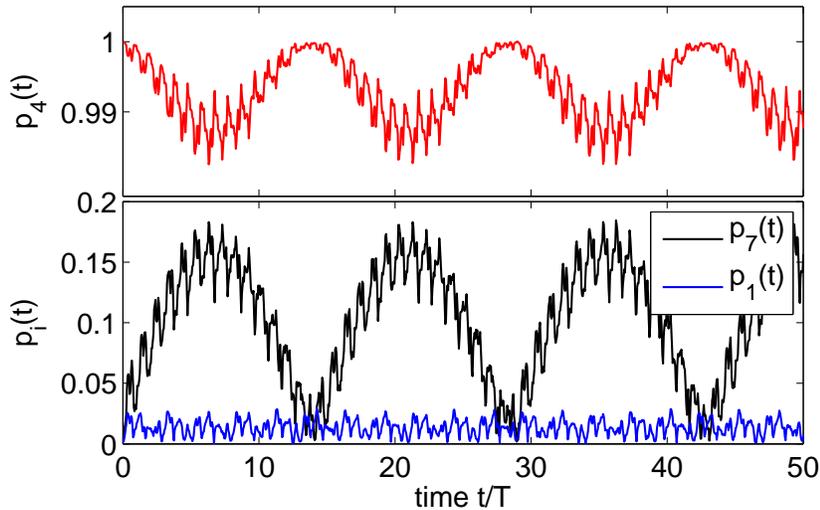}}
\caption{(Color online) Time-evolution of the projections $p_i(t)$ onto the instantaneous eigenstates when starting in the fourth state, i.e. $p_4(t)=1$ with a driving frequency of $\omega=5$, cf. \fref{fig:fig2}. Besides $p_4$, now $p_7$ gets considerably populated, note the large beating period $T_b>T$ of both coefficients. \label{fig:fig3}}
\end{figure}
In \fref{fig:fig2}, the evolution of the energy $E(t)$ is shown for a driving frequency of $\omega=5$ and two different initial states, namely the ground state (first state) and the third excited state (fourth state), i.e. $\Psi_0=\Psi_{0,1}^e(x,y)$ and $\Psi_0=\Psi_{2,1}^e(x,y)$, respectively. These are the first two states that have the same symmetry, namely $\pi_x=\pi_y=+1$, see \tref{tb:tab1}. When starting in the instantaneous ground state (at $t=0$), the wave function $\Psi(t)$  follows the motion of the elliptic boundary adiabatically. In other words, the projection $p_1(t)$ of $\Psi(t)$ on the instantaneous ground state is approximately one, $p_1(t)=|\braket{\Psi_{0,1}^e(t)|\Psi(t)}|\approx 1$. The periodic oscillations of the energy $E(t)$ are then simply present due to the fact that the energy of the instantaneous ground state changes periodically (approximately sinusoidal), cf. \fref{fig:fig1}. 

The situation is different when starting in the fourth state ($\Psi_0=\Psi_{2,1}^e(x,y)$). Even though $E(t)$ performs sinusoidal-like oscillations corresponding to the energy oscillations of the fourth  instantaneous eigenstate, there is a small modulation (beating) on top, with a period $T_b > T=2\pi/\omega$ much larger then the period of the applied driving law. This large-periodic modulation is  better visible in the time-evolution of the population coefficients $p_i(t)$, which are shown in \fref{fig:fig3}. Since $\Psi_0=\Psi_{2,1}^e(x,y)$, $p_4(0)=1$ and all the other $p_i(0)$ are zero. The fourth instantaneous eigenstate stays populated ($p_4(t)>0.98, \forall t$) throughout the whole time-evolution, however - besides small, fast oscillations with the period $T$ of the applied driving law - there is now a beating with a much larger period $T_b$. Correspondingly, other states get noticeably populated, primarily $p_7(t)$ and $p_1(t)$ are now greater than zero, with $p_7$ having the same large beating period $T_b$. Note that the instantaneous eigenstates corresponding to $p_1$, $p_4$ and $p_7$ share the same symmetry properties $\pi_x=\pi_y=+1$, cf. Table~\ref{tb:tab1}. 
\begin{figure}%[ht]
\centerline{\includegraphics[width=0.7\columnwidth]{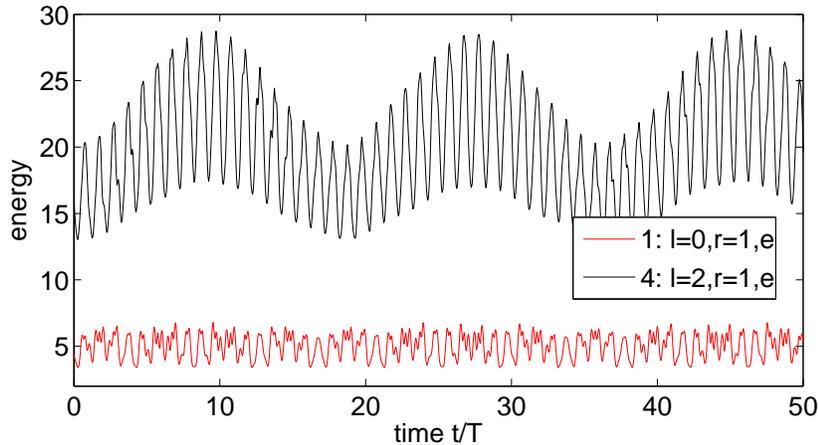}}
\caption{(Color online)  Evolution  $ E(t)$ for a driving frequency  $\omega=10$, when starting in the instantaneous ground state (first state) and in the fourth state. Whereas the fourth state shows a pronounced beating phenomenon, this is not the case for in the ground state. \label{fig:fig4}}
\end{figure}

\begin{figure}%[ht]
\centerline{\includegraphics[width=0.7\columnwidth]{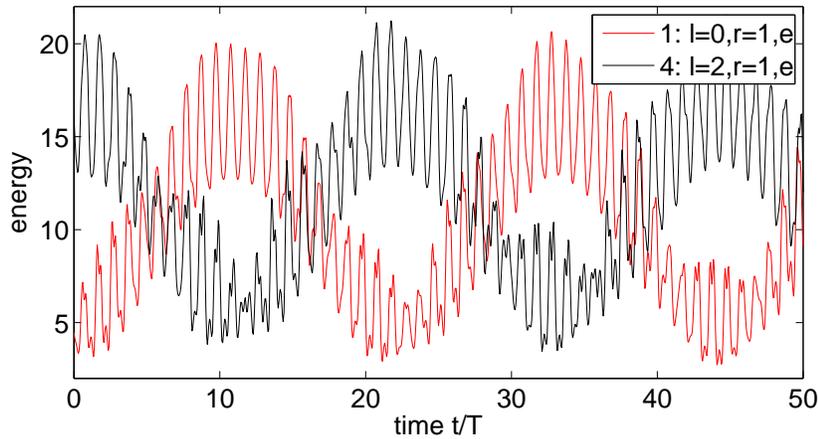}}
\caption{(Color online) Evolution  $ E(t)$ for a driving frequency of $\omega=12$, when starting in the instantaneous ground state (first state) and in the fourth state. Both curves show a pronounced beating phenomenon.  \label{fig:fig5}}
\end{figure}

When choosing other low energy eigenfunctions as an initial state, e.g. $\Psi_0=\Psi_{1,1}^e(x,y)$, $\Psi_{1,1}^o(x,y)$, $\Psi_{2,1}^o(x,y)$ or $\Psi_0=\Psi_{3,1}^e(x,y)$, the result is very much like the one when starting with $\Psi_0=\Psi_{0,1}^e(x,y)$, in the sense that the energy $E(t)$ performs periodic oscillations matching the sinusoidal-like oscillations of the corresponding instantaneous eigenstate. This behavior is also observed for smaller driving frequencies ($\omega<4$), now also when starting in the fourth state, i.e. $\Psi_0=\Psi_{2,1}^e(x,y)$.

\begin{figure}%[ht]
\centerline{\includegraphics[width=0.7\columnwidth]{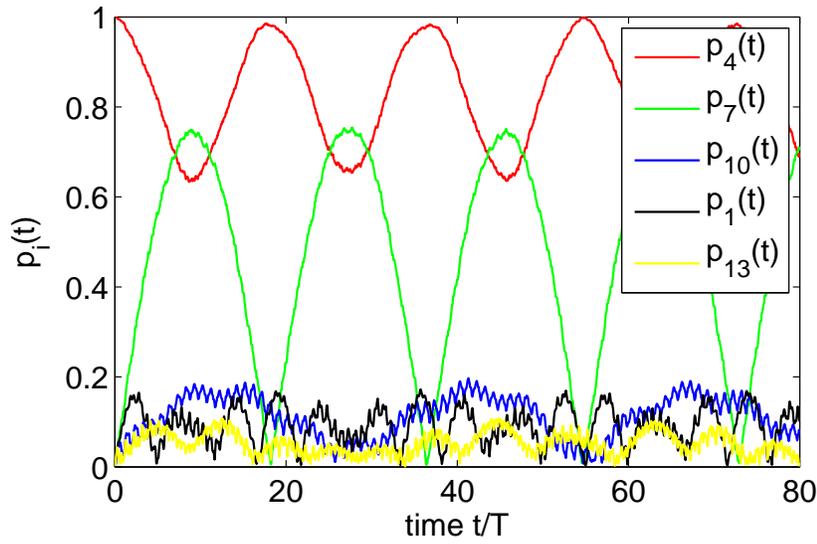}}
\caption{(Color online) Time-evolution of the coefficients $p_i(t)$ corresponding to the upper curve (state 4) of \fref{fig:fig4} ($\omega=10$). Again $p_4$ and $p_7$ dominate, however, now further states ($p_{10}, p_1, p_{13}$) contribute.  \label{fig:fig6}}
\end{figure}

For larger driving frequencies, the beating phenomenon becomes even more pronounced. In \fref{fig:fig4}, a driving frequency of $\omega=10$ is applied to $\Psi_0=\Psi_{0,1}^e(x,y)$ and $\Psi_0=\Psi_{2,1}^e(x,y)$. For $\Psi_0=\Psi_{0,1}^e(x,y)$ the resulting $E(t)$ shows besides the large amplitude beating some irregular fluctuations, however $E(t)$ stays tightly within a well-defined band. Now for $\Psi_0=\Psi_{2,1}^e(x,y)$ the beating is much more pronounced. Again, the period $T_b$ of the beating is much larger than the driving period $T$, note however that it is not the same beating period we obtained in the case of $\omega=5$, cf. \fref{fig:fig3}. Like in the case $\omega=5$, the time-evolution of the population coefficients, see \fref{fig:fig6}, reflects the beating phenomenon. Again (for $\Psi_0=\Psi_{2,1}^e(x,y)$), the coefficient $p_7$ is the one right behind $p_4$ that gets populated the most, note that at some times even $p_7>p_4$. Other states with noticeable population are $p_{10}$, $p_1$ and $p_{13}$, again these are all states with symmetries $\pi_x=\pi_y=+1$, cf. Table~\ref{tb:tab1}.

\begin{figure}%[ht]
\centerline{\includegraphics[width=0.7\columnwidth]{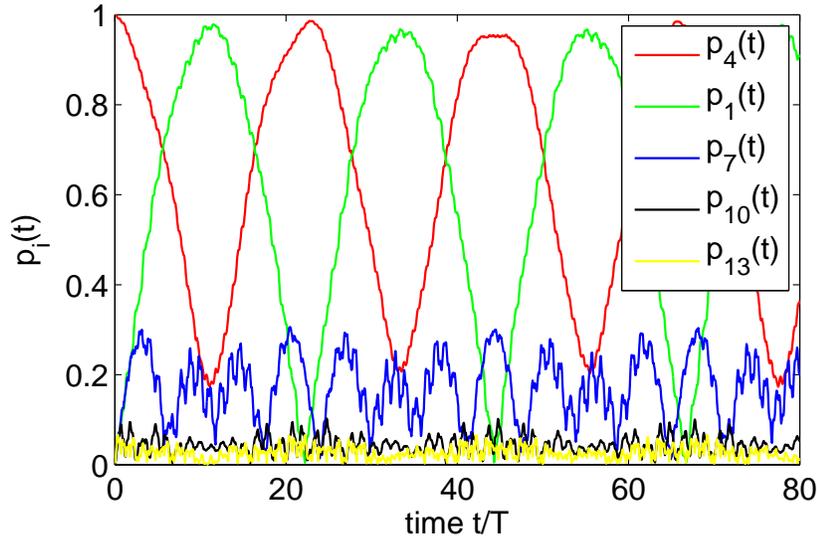}}
\caption{(Color online) Time-evolution of the coefficients $p_i(t)$ corresponding to  \fref{fig:fig5} ($\omega=12$), when starting in the fourth state. Unlike in the case $\omega=10$, besides $p_4$, now $p_1$  (instead of $p_7$) dominates.  \label{fig:fig7}}
\end{figure}

By increasing the driving frequency further to $\omega=12$, the beating phenomenon is now observed equally when starting in the instantaneous ground state ($\Psi_0=\Psi_{0,1}^e(x,y)$) and when starting in the fourth state ($\Psi_0=\Psi_{2,1}^e(x,y)$), see \fref{fig:fig5}. Unlike in the case $\omega=10$, the large-periodic modulation  (in the case $\Psi_0=\Psi_{2,1}^e(x,y)$) considerably lowers rather than rises the energy $E(t)$. Consequently, the state besides $p_4$ that gets substantially populated is now $p_1$ and not the higher energetic $p_7$, see \fref{fig:fig7}. Even more, $p_1$ reaches almost one, i.e. almost the whole population is periodically transferred between $p_4$ to $p_1$.     

\begin{figure}%[ht]
\centerline{\includegraphics[width=0.7\columnwidth]{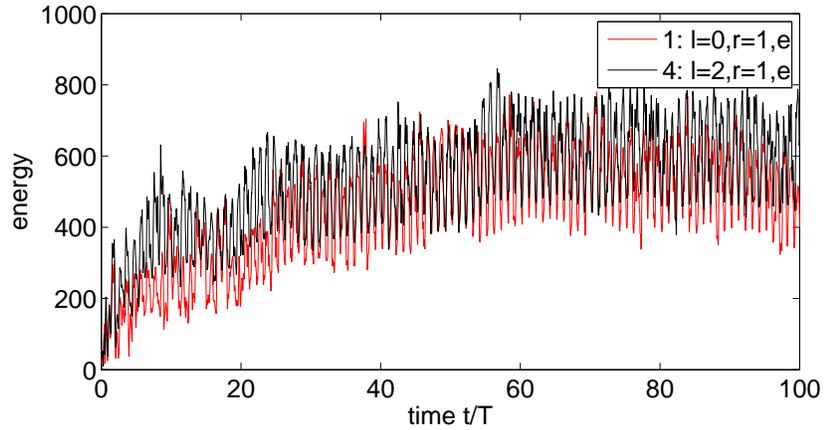}}
\caption{(Color online)  Evolution of the energy $ E(t)$ for a driving frequency $\omega=100$, when starting in the instantaneous ground state (first state) and in the fourth state. After roughly 60 driving periods, the energy $E(t)$ saturates.  \label{fig:fig8}}
\end{figure}

The beating phenomenon gets destroyed for sufficiently high driving frequencies, see \fref{fig:fig8}, where $E(t)$ for $\omega=100$ with $\Psi_0=\Psi_{0,1}^e(x,y)$ ($p_1(0)=1$) and $\Psi_0=\Psi_{2,1}^e(x,y)$ ($p_4(0)=1$) is shown. Now a considerable number of states with the same symmetry as the initial state get noticeably populated. Consequently, the energy does not stay in a narrow interval as in the case of small driving frequencies, but rather fluctuates irregularly over a wide range. More specific, $E(t)$ increases rapidly within the first few oscillations of the elliptical billiard, then slowly grows further until $t\approx 60\, T$ and saturates for $t>60\, T$. This means that the energy stays bounded, unlike in the corresponding classical system \cite{Lenz2009,Lenz2010},  there is no Fermi acceleration in the quantum version. This is due to dynamical localization: the wave function gets exponentially localized in momentum space, thereby stopping the energy diffusion.    

\begin{figure}%[ht]
\centerline{\includegraphics[width=0.7\columnwidth]{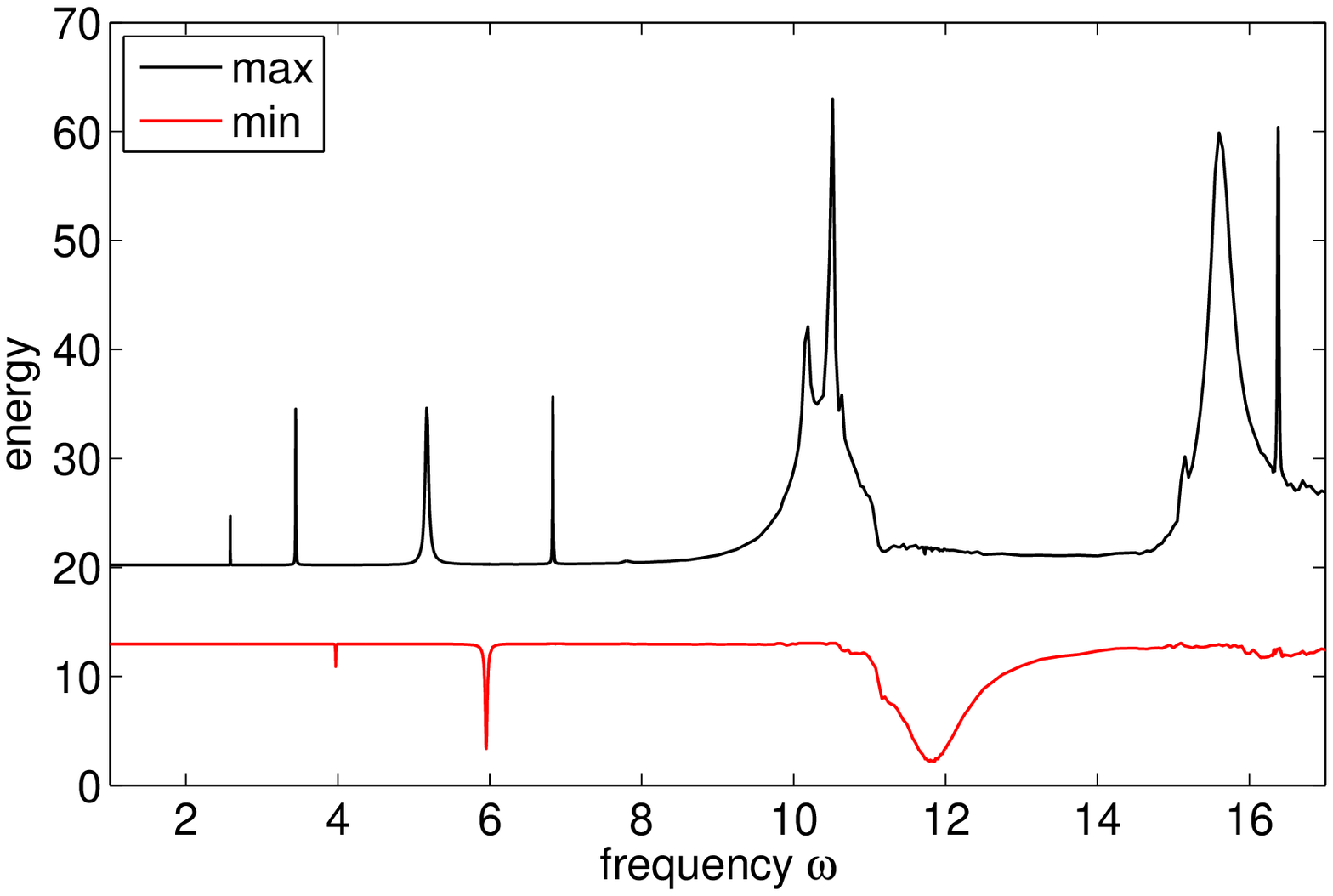}}
\caption{(Color online) Maximal and minimal energy reached when starting in the fourth state as a function of the driving frequency $\omega$. $E_{\max}$ and $E_{\min}$ show a series of resonances, which are related to a matching of the driving frequency $\omega$ to the mean energetic difference of instantaneous eigenstates, see \tref{tb:tab2}. \label{fig:fig9}}
\end{figure}

In general, when starting at $t=0$ with an instantaneous eigenstate with energy $E_i$, the maximal $E_{\max}= \max ( E(t)) $ and minimal energy $E_{\min}= \min( E(t)) $ of $\Psi(t)$ is a good indicator whether the time-evolution of $\Psi(t)$ is adiabatic or not. If the evolution of $\Psi(t)$ is adiabatic, $E_{\max,\min}$ are simply given by the corresponding eigenvalues of the instantaneous eigenstate, more specific, in this case ($\omega=1$) $E_{\max}=E_i(\zeta=3/4)$ and $E_{\min}=E_i(\zeta=1/4)$, cf. \fref{fig:fig1}. As soon as the time-evolution is not purely adiabatic, the beating phenomenon sets in and either $E_{\max}$ is raised, or $E_{\min}$ is lowered, or both. This can be seen in \fref{fig:fig9}, where $E_{\max}$ and $E_{\min}$ are shown as a function of the driving frequency $\omega$ when starting with the fourth state ($\Psi_0=\Psi_{0,1}^e(x,y)$). For small driving frequencies ($\omega<2.5$), $E_{\max}$ and $E_{\min}$ are given by the energies of the fourth instantaneous eigenstate, i.e. $E_{\max}=E_4(3/4)$ and $E_{\min}=E_4(1/4)$, respectively, which means that the time-evolution is adiabatic. For higher driving frequencies,  a series of resonances appear: $E_{\max}$ has sharp resonances at $\omega= 2.58,\,3.45,\,5.17,\,6.83$ and broad resonances with double peaks at $\omega=10.19,\,10.51$ and $\omega=15.6,\,16.4$. On the other hand, $E_{\min}$ shows considerably fewer resonances, sharp ones at $\omega=3.97,\,5.95$ and a broad one at $\omega=11.89$. The fact that $E_{\min}$ has considerably fewer resonances than $E_{\max}$ is immediately clear, since the only way that the energy, when starting in the fourth state ($p_4(0)=1$), can be lowered is to populate $p_1$, whereas many different energetically higher lying instantaneous states can be populated, leading to an increase of $E_{\max}$.    

To gain an understanding of the above presented results, in particular of the resonances of $E_{\max}$ and $E_{\min}$, we develop a Rabi-like model in the following section. 

\begin{table}
\caption{(color online) Mean (averaged of the phase $\zeta$) energy differences $E_{i\leftrightarrow j}$ together with $E_{i\leftrightarrow j}/2$, $E_{i\leftrightarrow j}/3$ and $E_{i\leftrightarrow j}/4$ for the first six states that share the same symmetry $\pi_x=\pi_y=+1$. The energy differences which are related to the resonances of \fref{fig:fig9} between just two states are shaded in gray, the ones related to the large double peak at $\omega\approx 10.5$ in yellow and the ones of the second double peak at $\omega\approx 15\dots\omega\approx16$ in green.}
\begin{center}
 \begin{tabular}{l|rrrr}\label{tb:tab2}
 & $E_{i \leftrightarrow j}$ & $E_{i \leftrightarrow j}/2$ & $E_{i \leftrightarrow j}/3$ & $E_{i \leftrightarrow j}/4$ \\
\hline
$E_{7 \leftrightarrow 10}$&\mc{1}{>{\columncolor{tcE}}r}{10.1833}&5.0917&3.3944&2.5458\\
$E_{4 \leftrightarrow 7}$&\mc{1}{>{\columncolor{tcE}}r}{10.3364}&\mc{1}{>{\columncolor{tcA}}r}{5.1682}&\mc{1}{>{\columncolor{tcA}}r}{3.4455}&\mc{1}{>{\columncolor{tcA}}r}{2.5841}\\
$E_{10 \leftrightarrow 13}$&\mc{1}{>{\columncolor{tcE}}r}{10.7058}&5.3529&3.5686&2.6764\\
$E_{1 \leftrightarrow 4}$&\mc{1}{>{\columncolor{tcA}}r}{11.9227}&\mc{1}{>{\columncolor{tcA}}r}{5.9613}&\mc{1}{>{\columncolor{tcA}}r}{3.9742}&2.9807\\
$E_{13 \leftrightarrow 18}$&17.5667&8.7833&5.8556&4.3917\\
$E_{4 \leftrightarrow 10}$&20.5198&\mc{1}{>{\columncolor{tcE}}r}{10.2599}&\mc{1}{>{\columncolor{tcA}}r}{6.8399}&5.1299\\
$E_{7 \leftrightarrow 13}$&20.8891&\mc{1}{>{\columncolor{tcE}}r}{10.4446}&6.9630&5.2223\\
$E_{1 \leftrightarrow 7}$&22.2591&11.1296&7.4197&5.5648\\
$E_{10 \leftrightarrow 18}$&28.2724&14.1362&9.4241&7.0681\\
$E_{4 \leftrightarrow 13}$&31.2255&\mc{1}{>{\columncolor{tcB}}r}{15.6128}&\mc{1}{>{\columncolor{tcE}}r}{10.4085}&7.8064\\
$E_{1 \leftrightarrow 10}$&32.4425&16.2212&\mc{1}{>{\columncolor{tcE}}r}{10.8142}&8.1106\\
$E_{7 \leftrightarrow 18}$&38.4558&19.2279&12.8186&9.6139\\
$E_{1 \leftrightarrow 13}$&43.1482&21.5741&14.3827&\mc{1}{>{\columncolor{tcE}}r}{10.7871}\\
$E_{4 \leftrightarrow 18}$&48.7922&24.3961&\mc{1}{>{\columncolor{tcB}}r}{16.2641}&12.1981\\
$E_{1 \leftrightarrow 18}$&60.7149&30.3575&20.2383&15.1787
\end{tabular}
\end{center}
\end{table}

\section{Interpretation and Rabi-like model}\label{ch:model}
The aim is now to  explain the resonances in \fref{fig:fig9} firstly by means of a population analysis of the instantaneous eigenstates and secondly via a Rabi-like model with time-periodic coupling.

Since the instantaneous eigenstates and the corresponding energies in the elliptical billiard are quantized, only certain states can be populated when driving with a specific frequency $\omega$. In particular, we expect an efficient transfer between two states, when their energy difference exactly matches the driving frequency, i.e. when $E_j-E_i=\omega$. As we saw earlier, the energy $E_i$ of an instantaneous eigenstate depends on the phase $\zeta$, it is thus useful to define the mean energy difference $E_{i\leftrightarrow j}$ between two instantaneous eigenstates as 
\begin{equation}
 E_{i\leftrightarrow j}:= \langle |E_i(\zeta)-E_j(\zeta)|\rangle_\zeta=\int_0^{1}|E_i(\zeta)-E_j(\zeta)|d\zeta.
\end{equation}
These mean differences together with $E_{i\leftrightarrow j}/2$, $E_{i\leftrightarrow j}/3$ and $E_{i\leftrightarrow j}/4$ are shown in \tref{tb:tab2} for the first six states that share the same symmetry $\pi_x=\pi_y=+1$. Now if we start in the fourth state ($p_4(0)=1$), cf. \fref{fig:fig9}, the only way to lower the energy is to populate the ground state ($p_1$). Thus, when $E_{1\leftrightarrow4}\approx \omega$, we expect a significant population of the ground state and consequently a decrease of $E_{\min}$. From \tref{tb:tab2} $E_{1\leftrightarrow4}=11.92$ and in \fref{fig:fig9}, there is a broad resonance around $\omega=11.89$. It is not surprising that the resonance is rather broad, since $E_{1\leftrightarrow4}$ is just the mean energy difference, actually the difference between the first and the forth state varies between $\max (E_4(\zeta)-E_1(\zeta))=14.6$ and $\min (E_4(\zeta)-E_1(\zeta))=9.6$, so transitions between the two states should be possible over a broad range around $\omega=11.89$ and this is exactly what we observe. 

The transition between the first and the fourth state is a one photon transition in the sense that $\omega=E_{1\leftrightarrow4}$, so one time the driving frequency matches the energy difference, yielding a broad resonance. On the other hand, there are also multi-photon transitions possible, such that $n\cdot \omega=E_{1\leftrightarrow4}, \, n=2,3,\dots$. Since the efficiency of multi-photon processes drastically reduces with the number of photons involved, these higher-order resonances should be much sharper and are naturally harder to detect. The sharp resonances of $E_{\min}$ at $\omega=5.95$ and $\omega=3.97$ are the two and three photon resonances of $E_{1\leftrightarrow4}$, respectively, and the numbers are in excellent agreement with the corresponding values of \tref{tb:tab2}.  

In a similar fashion, the sharp resonances of $E_{\max}$ can be explained: $\omega= 2.58,\,3.45,\,5.17$ correspond to the four, three, and two photon transitions of $E_{4\leftrightarrow 7}$, respectively and $\omega=6.83$ to the three photon transition of $E_{4\leftrightarrow 10}$, again in excellent agreement with the results of \tref{tb:tab2}. Note that a population analysis of all the resonances confirms the so far presented considerations, e.g. at $\omega=6.83$ besides $p_4$, $p_{10}$ gets populated, whereas at $\omega=5.17$, besides $p_4$ now this is the case for $p_7$.

Applied to the broad double peaks of  $E_{\max}$, a population analysis yields the following: at $\omega=10.19$ a mixture of $p_4$, $p_7$ and $p_{10}$ is present (also $p_{13}$, $p_1$ and $p_{20}$ are noticeably, however less than the first three ones populated), whereas at $\omega=10.51$ mostly the states $p_7$, $p_4$, $p_{20}$, $p_{13}$, $p_{10}$ and $p_{1}$ contribute. The yellow shielded fields of \tref{tb:tab2} indicate that there is a multitude of one- to four-photon transitions in this frequency regime between exactly these states that according to the population analysis contribute to the resonance. Note that there are now also processes of the kind $p_4\rightarrow p_7 \rightarrow p_{10}$ possible, since the corresponding energy differences $E_{4\leftrightarrow7}$ and $E_{7\leftrightarrow10}$ are comparable. In contrast to the just described resonance, the broad double peak at $\omega=15.6$ and $\omega=16.4$ is mainly due to transitions between just  two states. The corresponding population analysis yields $p_4$ and $p_{13}$ for $\omega=15.6$ and $p_4$ and $p_{18}$ for $\omega=16.4$, again in excellent agreement with the two-photon transition of $E_{4\leftrightarrow13}$ and the three-photon transition of $E_{4\leftrightarrow18}$, see the greenish fields of \tref{tb:tab2}.   

For all the resonances, the population analysis additionally shows that the beating period $T_b$ increases dramatically when approaching a resonance, i.e. the population transfer from one instantaneous eigenstate to another becomes slower and slower the closer the driving frequency gets to the corresponding resonance. 

\begin{figure}%[ht]
\centerline{\includegraphics[width=0.7\columnwidth]{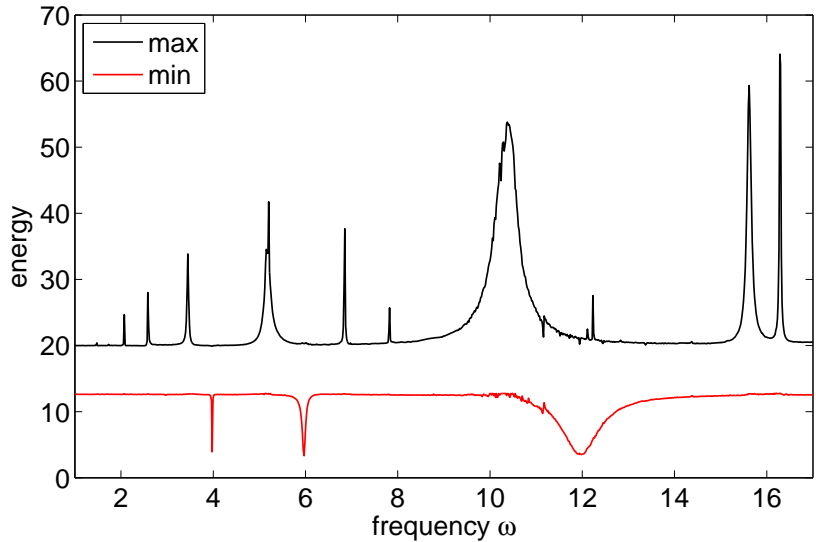}}
\caption{(Color online) Maximal and minimal energy reached when starting in fourth state as a function of the driving frequency $\omega$ now for the Rabi-like model. There is very good  agreement with the full numerical simulations, see \fref{fig:fig9}. \label{fig:fig10}}
\end{figure}

In the discussion of the resonances, we already used the projection onto the instantaneous eigenstates to analyze the results.  With this picture in mind (see e.g. \fref{fig:fig3}, where just two states are considerably populated), a very crude simplification is to interpret the driven elliptical billiard as a two-level system (two instantaneous eigenstates), where the two levels are coupled by a periodic driving with an unknown coupling strength $a$. Without the driving, the Hamiltonian of the unperturbed system  in a suitable basis is then simply given by 
\begin{equation*}
 \mathcal H^0 = \left(
  \begin{array}{cc}
   E_1 & 0  \\
    0 & E_2
   \end{array} \right), 
\end{equation*}
where $E_1$ and $E_2$ are the energies of the two eigenstates $\Psi_1^0$ and $\Psi_2^0$, with time-evolution
\begin{equation*}
 \Psi_n^0(t)=\Psi_n^0 e^{-iE_nt/\hbar}, \quad n=1,2.
\end{equation*}
Next, we take into account the driving of the system, which leads to a periodic coupling of the two states. In the Hamiltonian, this leads to off-diagonal elements of the form $a\sin (\omega t)$. However, as  can be seen in \fref{fig:fig1}, the energy itself of the instantaneous eigenstates depends on time. To this end, we introduce  time-dependent energy shifts $\triangle E_i(t)$ on the diagonal. In a good approximation, the  $\triangle E_i(t)$ are harmonic functions, which we obtain by fitting sine functions to the curves of \fref{fig:fig1}. Overall, the model Hamiltonian is then
\begin{equation*}
 \mathcal H(t)=\mathcal H^0 + \mathcal H^1(t),
\end{equation*}
where
\begin{equation*}
\mathcal H^1(t) = \left(
  \begin{array}{cc}
   \triangle E_1(t) & a \sin (\omega t)  \\
    a \sin (\omega t) & \triangle E_2(t)
   \end{array} \right).
\end{equation*}
The standard ansatz for the time-dependent wave function $\Psi(t)$ is a superposition with time-dependent coefficients of the unperturbed wave functions $\Psi^0_i(t)$, i.e.
\begin{equation*}
  \Psi(t)= c_1(t)\Psi_1^0(t)+c_2(t)\Psi_2^0(t).
\end{equation*}
Plugging this ansatz into the time-dependent Schr\"odinger equation leads to a system of ordinary differential equations for the coefficients
\numparts\label{eq:rabi_coefficients}
\begin{eqnarray}
 \dot c_1(t)&= \frac{1}{i\hbar} \left[c_2(t) a \sin (\omega t) e^{i(E_1-E_2)t/\hbar}+c_1(t) \triangle E_1(t)  \right]\\
\dot c_2(t)&= \frac{1}{i\hbar} \left[c_1(t) a \sin (\omega t) e^{i(E_2-E_1)t/\hbar}+c_2(t) \triangle E_2(t)  \right].
\end{eqnarray}
\endnumparts
Note that at this point, usually (e.g. in atomic systems) the rotating wave approximation is applied, however, in our case this would lead to incorrect results, so we have to solve the full system given by Eqs.~\eref{eq:rabi_coefficients}, which is only possible numerically. 

Not surprisingly, this simple 2-level model does not lead to correct results, e.g. the 2-level model cannot reproduce all the resonances of \fref{fig:fig9}. The obvious reason is, as the population analysis shows, that already for driving frequencies in the range shown in \fref{fig:fig9} ($\omega<17$) more than just two states are involved. The straightforward generalization to n-levels yields:
\begin{equation*}
 \mathcal H(t)=\mathcal H^0 + \mathcal H^1(t)
\end{equation*}
where now $\mathcal H^1(t)$ reads
\begin{equation*}
\mathcal H^1(t) = \left(
  \begin{array}{cccc}
   \triangle E_1(t) &  a \sin \omega t & \dots &   \\
    a\sin \omega t & \triangle E_2(t) & & \\
  \vdots & & \ddots & \\
     & & & \triangle E_n(t)
   \end{array} \right),
\end{equation*}
or in other words $ \mathcal H_{ii}^1(t)=\triangle E_i(t)$ and for the off-diagonal elements we have $ \mathcal H_{ij}^1(t)=a \sin(\omega t)$. It is reasonable to assume that the coupling strength $a$ is equal between all the levels, leaving us with just one free parameter in the model. Again, the time-evolution of the unperturbed states is given by
\begin{equation*}
 \mathcal H^0 \Psi_n^0= E_n \Psi_n^0 \, \Rightarrow \, \Psi_n^0(t)=\Psi_n^0 e^{-iE_nt/\hbar}
\end{equation*}
and the superposition ansatz reads 
\begin{equation*}
  \Psi(t)= \sum_n c_n(t)\Psi_n^0(t),
\end{equation*}
yielding a set of coupled linear differential equations for the coefficients, which has to be solved numerically:
\begin{equation*}
 \dot c_i(t)= \frac{1}{i\hbar}\sum_j c_j(t) \mathcal H_{ij}^1(t) e^{i(E_i-E_j)t/\hbar}.
\end{equation*}
With this Rabi-like model, we calculate the time evolution of $\Psi(t)$, with the same initial conditions as in the discussion of the resonances, i.e. we start in the fourth state, $c_4(0)=1$. We take into account the first six states that can couple, i.e. the states with symmetries $\pi_x=\pi_y=+1$, namely $1$, $4$, $7$, $10$, $13$ and $18$, so we utilize a six-level Rabi-like model. Just like for the full numerical calculations, we compute $E_{\max}$ and $E_{\min}$ as a function of the applied driving frequency $\omega$. The result is shown in \fref{fig:fig10}. Overall, there is very good agreement with the results of the full simulations, cf. \fref{fig:fig9}. In particular, all resonances of \fref{fig:fig9} are reproduced. The additional resonances that appear in the Rabi-like model are four-, five- and even six-photon processes: $\omega=2.07$ corresponds to the five-photon transition of $E_{4\leftrightarrow7}$,  $\omega=7.81$ to the four-photon transition of $E_{4\leftrightarrow13}$ and $\omega=12.21$ to the four-photon transition of $E_{4\leftrightarrow18}$. The resonance at $\omega=5.17$ is superimposed with a very sharp resonance at $\omega=5.20$. This is the six-photon transition of $E_{4\leftrightarrow13}$, additionally, we suppose that this resonance is enhanced by the combination of the two-photon transition $E_{4\leftrightarrow7}$ with the four-photon transition $E_{7\leftrightarrow13}$, again cf. \tref{tb:tab2}. 

The additional resonances are extremely sharp and this is one reason why they do not appear in \fref{fig:fig9}. A much finer frequency scan would be necessary in the case of the full numerical simulations. Secondly, as already mentioned, the population transfer between two states becomes extremely slow when approaching a resonance (the beating period $T_b$ increases), i.e. the simulations have to be carried out over longer times in order to detect the very sharp resonances. 

The double peak at $\omega=10.19,\,10.51$ of \fref{fig:fig9} appears in the Rabi-like model as a single, broad resonance, see \fref{fig:fig10}. The reason is that the time-dependent energy shifts $\triangle E_i(t)$ of the model are obtained by fitting sine functions to the energy levels $E_i(\zeta)$ of \fref{fig:fig1}. This fitting procedure does preserve the mean distances between the energy levels, in particular it reproduces the correct $E_{i\leftrightarrow j}$, however, the minima of $\triangle E_i(t)$ are lowered and the maxima of $\triangle E_i(t)$ are raised compared to $E_i(\zeta)$. As a consequence, the resonances are broadened in the Rabi-like model, such that the mentioned double-peak merges, so they cannot be resolved within the model. Note that when fitting the $E_i(\zeta)$ in a different fashion, such that the minima and maxima are preserved, the double peak does appear in the model, however this alternative fitting procedure changes the $E_{i\leftrightarrow j}$, so all the resonances are shifted.

The fact that the beating period $T_b$ increases drastically when approaching a resonance is also reproduced by the Rabi-like model, however, unlike for the appearance of the resonances itself, we do not have an intuitive explanation for these beatings. Nevertheless, the success of the Rabi-like model shows that we have identified the essential ingredients to explain the time-evolution of a wave function, when starting in an eigenstate, in the driven elliptical billiard.    

\section{Conclusion and Outlook}\label{ch:outlook}
In this work, we have developed a numerical method to propagate an arbitrary initial state in the time-dependent elliptical billiard characterized by time-dependent boundary conditions. The method is based on a series of transformations of the original Hamiltonian, removing the time-dependent boundary conditions. Subsequently expanding the wave function in a suitable basis, we obtain a large system of coupled ordinary differential equations for the time-dependent expansion coefficients which can be solved by standard techniques. We then propagate eigenstates of the static system and investigate the time-evolution of their energy in the driven billiard. Unlike in the corresponding classical system, the energy does not grow unboundedly  but rather saturates, even for large driving frequencies, i.e. there is no Fermi acceleration in the quantum billiard. The  maximal and minimal energy that is reached as a function of the applied driving frequency show a series of resonances. We analyze these resonances by projecting the wave function onto the instantaneous eigenstates, yielding the spectral composition of the states. At these resonances, population is periodically, with a period $T_b$, transferred between the initial state and another one, either increasing or decreasing the energy of the wave function. This analysis shows that the resonances appear exactly when the driving frequency $\omega$ matches the mean energetic difference between the involved instantaneous states (one photon transition) or multiples of the driving frequency ($n$-photon transitions, which result in very sharp resonances). The beating period $T_b$ of the population transfer is much larger than the period $T=2\pi/\omega$ of the driving and increases dramatically when approaching a resonance. To confirm our intuitive picture describing the appearance of the resonances, we  develop a few-level Rabi-like model with time-periodic couplings, and time-dependent energy shifts on the diagonal, resembling the phase dependence of the energies of the instantaneous eigenstates. This model  reproduces nicely the essential results of our full numerical simulations, in particular the mentioned resonances and the occurrence of the beating period $T_b$.

So far, we investigated the breathing mode of the elliptical billiard, i.e. the time-dependence of the semi-major and semi-minor axes is given by $a(t)=a_0+c\sin(\omega t)$ and $b(t)=b_0+c\sin(\omega t)$, respectively. A promising alternative would be to use a time-law obeying $a(t)b(t)=const.$, since this preserves the area of the ellipse. As a result, the instantaneous energy levels exhibit a multitude of crossings, naturally posing the question whether $a(t)$ could be adjusted in such a way that the momentum diffusion does not stop, yielding  quantum Fermi acceleration in the driven elliptical billiard. 

\appendix
\section{Details of the ODE system}\label{ch:appendix}
The $d^{(i)}_{nmn'}(t)$ of the ODE system, cf. Eq.~\eref{eq7:ode_short}, are given by
\begin{eqnarray}\label{eq7:d1}
d^{(1)}_{nmn'}(t)=
\frac{1}{J_{m+1}(k_{m,n})J_{m+1}(k_{m,n'})}\bigg[\left(-\frac{\hbar^2}{\mu a^2(t)}-\frac{\hbar^2}{\mu b^2(t)}\right)\times \nonumber \\
\bigg(\frac{k_{m,n'}^2}{8}(L^1_{nmn'}-2L^2_{nmn'}+L^3_{nmn'})
-\frac{k_{m,n'}(m-1)}{4}L^{10}_{nmn'}\nonumber \\-\frac{k_{m,n'}(m+1)}{4}L^{11}_{nmn'}\bigg)
+\frac{\mu
L^{16}_{nmn'}}{2}\left(a(t)\ddot a(t)+b(t)\ddot b(t)\right) \bigg],
\end{eqnarray}
\begin{eqnarray}
d^{(2)}_{nmn'}(t)= \frac{1}{J_{m+1}(k_{m,n})J_{m-1}(k_{m-2,n'})}\bigg[\left(\frac{\hbar^2}{\mu a^2(t)}-\frac{\hbar^2}{\mu b^2(t)}\right)\times \nonumber \\
\bigg(\frac{k_{m-2,n'}^2}{16}(L^7_{nmn'}-2L^8_{nmn'}+L^9_{nmn'})
+\frac{3k_{m-2,n'}(m-1)}{8}L^{13}_{nmn'}\nonumber \\-\frac{k_{m-2,n'}(m-3)}{8}L^{12}_{nmn'}
\bigg)
+ \frac{\mu L^{18}_{nmn'}}{4}\left(a(t)\ddot a(t)-b(t)\ddot
b(t)\right) \bigg],
\end{eqnarray}
\begin{eqnarray}\label{eq7:d3}
d^{(2)}_{nmn'}(t)= \frac{1}{J_{m+1}(k_{m,n})J_{m+3}(k_{m+2,n'})}\bigg[\left(\frac{\hbar^2}{\mu a^2(t)}-\frac{\hbar^2}{\mu b^2(t)}\right)\times \nonumber \\
\bigg(\frac{k_{m+2,n'}^2}{16}(L^4_{nmn'}-2L^5_{nmn'}+L^6_{nmn'})
+\frac{3k_{m+2,n'}(m+1)}{8}L^{14}_{nmn'}\nonumber \\
-\frac{k_{m+2,n'}(m+3)}{8}L^{15}_{nmn'}\bigg)+ \frac{\mu L^{17}_{nmn'}}{4}\left(a(t)\ddot a(t)-b(t)\ddot
b(t)\right) \bigg],
\end{eqnarray}
where $a(t)$ and $b(t)$ are determined by the driving law of the elliptical billiard and the $L^i_{nmn'}$ are defined as the following integrals over products of Bessel functions (for details of the derivation see \cite{LenzThesis}), where we use
\begin{equation}
 I_{nmn'}(q,p,s):=  \int_0^1 J_m(k_{m,n}r) J_{m+p}(k_{m+q,n'}r)r^sdr
\end{equation}
and in the following we will omit the indeces $nmn'$ for readability:
\begin{eqnarray*}
&L^1 =I(-2,0,1) \qquad &L^2=I(0,0,1)\\
&L^3 =I(2,0,1)  &L^4=I(0,2,1)\\
&L^5 =I(2,2,1)  &L^6=I(4,2,1)\\
&L^7 =I(-4,-2,1)  &L^8=I(-2,-2,1)\\
&L^9 =I(0,-2,1)  &L^{10}=I(-1,0,0)\\
&L^{11} =I(1,0,0)  &L^{12}=I(-3,-2,0)\\
&L^{13} =I(-1,-2,0)  \qquad &L^{14}=I(1,2,0)\\
&L^{15} =I(3,3,0)  &L^{16}=I(0,0,3)\\
&L^{17} =I(2,2,3)  &L^{18}=I(-2,-2,3)
\end{eqnarray*}
The $d^{(i)}_{nmn'}(t)$ can be decomposed into time-independent parts $f^i_{nmn'}$ and simple time-dependent functions $g_i(t)$ in the following way:
%\numparts
 \begin{eqnarray}
 d^{(1)}_{nmn'}(t)&= g^{(1)}(t)f^{(1)}_{nmn'}+g^{(2)}(t)f^{(2)}_{nmn'}\\
 d^{(1)}_{nmn'}(t)&= g^{(3)}(t)f^{(3)}_{nmn'}+g^{(4)}(t)f^{(4)}_{nmn'}\\
 d^{(1)}_{nmn'}(t)&= g^{(3)}(t)f^{(5)}_{nmn'}+g^{(4)}(t)f^{(6)}_{nmn'}
\end{eqnarray}
%\endnumparts
with
%\numparts
 \begin{eqnarray}
g^{(1)}(t)&=-\frac{\hbar^2}{\mu a^2(t)}-\frac{\hbar^2}{\mu b^2(t)}\\
g^{(2)}(t)&=\mu \left (a(t)\ddot a(t)+b(t)\ddot b(t)\right )\\
g^{(3)}(t)&=\frac{\hbar^2}{\mu a^2(t)}-\frac{\hbar^2}{\mu b^2(t)}\\
g^{(4)}(t)&=\mu  \left ( (a(t)\ddot a(t)-b(t)\ddot b(t) \right)
\end{eqnarray}
%\endnumparts
and
 \begin{eqnarray}
 f^{(1)}_{nmn'}:=  \frac{k_{m,n'}^2(L^1_{nmn'}-2L^2_{nmn'}+L^3_{nmn'})-2k_{m,n'}(m-1)L^{10}_{nmn'}}{8J_{m+1}(k_{m,n})J_{m+1}(k_{n'm})}\nonumber \\
-\frac{2k_{m,n'}(m+1)L^{11}_{nmn'}}{8J_{m+1}(k_{m,n})J_{m+1}(k_{n'm})}\nonumber \\
f^{(2)}_{nmn'}:= \frac{L^{16}_{nmn'}}{2J_{m+1}(k_{m,n})J_{m+1}(k_{n'm})}\nonumber \\
f^{(3)}_{nmn'}:=\frac{k_{m-2,n'}^2(L^7_{nmn'}-2L^8_{nmn'}+L^9_{nmn'})+6k_{m-2,n'}(m-1)L^{13}_{nmn'}}{16J_{m+1}(k_{m,n})J_{m-1}(k_{m-2,n'})}\nonumber \\
-\frac{2k_{m-2,n'}(m-3)L^{12}_{nmn'}}{16J_{m+1}(k_{m,n})J_{m-1}(k_{m-2,n'})}\nonumber \\
f^{(4)}_{nmn'}:=\frac{L^{18}_{nmn'} } {4J_{m+1}(k_{m,n})J_{m-1}(k_{m-2,n'})}\nonumber \\
f^{(5)}_{nmn'}:=\frac{k_{m+2,n'}^2(L^4_{nmn'}-2L^5_{nmn'}+L^6_{nmn'})+6k_{m+2,n'}(m+1)L^{14}_{nmn'}}{16J_{m+1}(k_{m,n})J_{m+3}(k_{m+2,n'})}\nonumber \\
-\frac{2k_{m+2,n'}(m+3)L^{15}_{nmn'}}{16J_{m+1}(k_{m,n})J_{m+3}(k_{m+2,n'})}\nonumber \\
f^{(6)}_{nmn'}:=\frac{L^{17}_{nmn'}}{4J_{m+1}(k_{m,n})J_{m+3}(k_{m+2,n'})}.
\end{eqnarray}

\section*{References}
\bibliographystyle{unsrt}
%\bibliography{thesis}

\end{document}